%% file: main.tex
\pdfoutput=1
\documentclass[12pt,a4paper]{article}
\usepackage{ifthen} % for conditional statements
\newboolean{pdflatex}
\setboolean{pdflatex}{true} % False for eps figures 

\newboolean{articletitles}
\setboolean{articletitles}{true} % False removes titles in references

\newboolean{uprightparticles}
\setboolean{uprightparticles}{false} %True for upright particle symbols

\newboolean{inbibliography}
\setboolean{inbibliography}{false} %True once you enter the bibliography

\input{preamble}
\input{mydefine}

\usepackage{longtable} % only for template; not usually to be used in PAPERs

\begin{document}

%%%%%%%%%%%%%%%%%%%%%%%%%
%%%%% Title     %%%%%%%%%
%%%%%%%%%%%%%%%%%%%%%%%%%
\renewcommand{\thefootnote}{\fnsymbol{footnote}}
\setcounter{footnote}{1}

% %%%%%%% CHOOSE TITLE PAGE--------
%\onecolumn
% \input{title-LHCb-ANA}
%\input{title-LHCb-CONF}
\input{title-LHCb-PAPER}

%\twocolumn
% %%%%%%%%%%%%% ---------

\renewcommand{\thefootnote}{\arabic{footnote}}
\setcounter{footnote}{0}

%%%%%%%%%%%%%%%%%%%%%%%%%%%%%%%%
%%%%%  Table of Content   %%%%%%
%%%%%%%%%%%%%%%%%%%%%%%%%%%%%%%%
%%%% Uncomment next 2 lines if desired
%\tableofcontents
%\cleardoublepage

%%%%%%%%%%%%%%%%%%%%%%%%%
%%%%% Main text %%%%%%%%%
%%%%%%%%%%%%%%%%%%%%%%%%%

\pagestyle{plain} % restore page numbers for the main text
\setcounter{page}{1}
\pagenumbering{arabic}

%% Uncomment during review phase. 
%% Comment before a final submission.
%\linenumbers

\noindent
In the Standard Model of particle physics the $B_c$ meson family is unique because it contains two different heavy flavor quarks, charm and beauty. 
The ground state of the $B_c$ meson family has a rich set of decay modes since either constituent quark can decay with the other as a spectator, or they can annihilate to a virtual $\W$ boson. 
The search for new $\Bcp$ decay channels\footnote{Charge conjugation is implied throughout the paper.} and precise measurements of their branching fractions can improve the understanding of Quantum Chromodynamics~(QCD) and can test various effective models. 
Many properties of the $\Bcp$ meson have been investigated by the $\lhcb$ experiment:
the $\Bcp$ mass, lifetime and production rate have been measured~\cite{LHCb-PAPER-2012-028,LHCb-PAPER-2013-010,LHCb-PAPER-2014-039,LHCb-PAPER-2013-063,LHCb-PAPER-2014-060,LHCb-PAPER-2014-050},
while several new decay channels have been observed~\cite{LHCb-PAPER-2011-044,LHCb-PAPER-2012-054,LHCb-PAPER-2013-010,LHCb-PAPER-2013-021,LHCb-PAPER-2013-044,LHCb-PAPER-2013-047,LHCb-PAPER-2014-009,LHCb-PAPER-2014-025,LHCb-PAPER-2014-039}.
The observation of the $\Bcp \to \psitwos \pi^+$ decay was made with $pp$ collision data at a center-of-mass energy of $\sqs=7\tev$, corresponding to an integrated luminosity of $1.0\invfb$~\cite{LHCb-PAPER-2012-054}. 
The ratio of the branching fraction of the $\Bcp \to \psitwos\pip$ decay with respect to that of the $\Bcp \to \jpsi\pip$ decay, defined as
\begin{equation}
\RB\equiv\frac{\BR(\Bcp \to \psitwos\pi^+)}{\BR(\Bcp \to \jpsi\pi^+)},
\end{equation}
was measured to be $\OldRes$.
The first uncertainty is statistical, the second is systematic, 
and the third is due to the uncertainties on the branching fractions of the $\JpsiToMuMu$ and $\PsiToMuMu$ decays.
The statistical uncertainty is dominant.
Several theoretical predictions for $\RB$ based on different effective models~\cite{Ebert:2003cn,Chang:1992pt,Liu:1997hr,Colangelo:1999zn,Qiao:2012hp,Rui:2015iia} exist, and vary between $0.07$ and $0.29$.

The analysis presented here updates the previous LHCb measurement of $\RB$~\cite{LHCb-PAPER-2012-054},
using the full $pp$ collision data collected by $\lhcb$ in 2011 and 2012 at $\sqs=7\tev$ and $8\tev$ respectively, corresponding to an integrated luminosity of $3\invfb$.
Due to the increased data sample and an improved analysis method, the statistical uncertainty is reduced by half, 
allowing a more powerful test of the theories.

The \lhcb detector~\cite{Alves:2008zz,LHCb-DP-2014-002} is a single-arm forward spectrometer covering the \mbox{pseudorapidity} range $2<\eta <5$,
and is designed for the study of particles containing \bquark or \cquark quarks.
The detector includes a high-precision tracking system consisting of a silicon-strip vertex detector surrounding the $pp$ interaction region~\cite{LHCb-DP-2014-001}, 
a large-area silicon-strip detector located upstream of a dipole magnet with a bending power of about $4{\rm\,Tm}$, 
and three stations of silicon-strip detectors and straw drift tubes~\cite{LHCb-DP-2013-003} placed downstream of the magnet.
The tracking system provides a measurement of momentum, \ptot, of charged particles with a relative uncertainty that varies from 0.5\% at low momentum to 1.0\% at 200\gevc.
The minimum distance of a track to a primary vertex~(PV), the impact parameter, is measured with a resolution of $(15+29/\pt)\mum$,
where \pt is the component of the track momentum transverse to the beam, in \gevc.
Different types of charged hadrons are distinguished using information from two ring-imaging Cherenkov detectors~\cite{LHCb-DP-2012-003}. 
Photons, electrons and hadrons are identified by a calorimeter system consisting of scintillating-pad and preshower detectors, an electromagnetic calorimeter and a hadronic calorimeter. 
Muons are identified by a system composed of alternating layers of iron and multiwire proportional chambers~\cite{LHCb-DP-2012-002}.
The online event selection is performed by a trigger~\cite{LHCb-DP-2012-004}, 
which consists of a hardware stage, based on information from the calorimeter and muon systems, 
followed by a software stage, which applies a full event reconstruction.

In the $\Bcp \to \jpsi\pip$ and $\Bcp \to \psitwos\pip$ decay channels, 
the $\jpsi$ and $\psitwos$ mesons are reconstructed through their decays into two muons.
At least one muon with high \pt is required in the hardware trigger.
The software trigger requires a charged particle with $\pt>1.7\gevc$, or $\pt>1\gevc$ if identified as a muon; 
alternatively a dimuon trigger requires two oppositely charged muons with $\pt>500\mevc$, and the invariant mass of the muon pair greater than $2.95\gevcc$.

Further offline selections require a good quality muon track with $\pt>550\mevc$, 
a good quality vertex for the reconstructed $\jpsi$ or $\psitwos$ candidate, 
and the reconstructed $\jpsi$ and $\psitwos$ masses to be 
within $\pm 100\mevcc$ of their known values~\cite{PDG2014}. 
The mass resolution for both resonances is $14\mevcc$.
The muon track pair and the pion track are required to be inconsistent with originating from a PV. 
The pion track is required to be of good quality, and to have a $\pt$ greater than $500\mevc$. 
The particle identification~(PID) information for pions is used to reduce the contamination from kaons and protons. 
The $\Bcp$ candidate is required to have a good quality vertex and a reconstructed mass within $\pm500\mevcc$ of its known mass~\cite{PDG2014},
which corresponds to more than ten times the mass resolution.
To further separate signal from background a boosted decision tree~(BDT) selection using the AdaBoost algorithm~\cite{Breiman,AdaBoost} is applied.
The selection uses more input variables and is more sophisticated compared to the previous analysis~\cite{LHCb-PAPER-2012-054}.

Simulated samples are generated to study the behaviour of signal events. 
The $\Bcp$ signals are generated with a dedicated generator $\bcvegpy$ \cite{Chang:2003cq,Chang:2005hq} through the dominant hard sub-process $gg \to \Bcp + \bquark + \cquarkbar$. 
The fragmentation and hadronisation processes are simulated with $\pythia$ \cite{Sjostrand:2006za,LHCb-PROC-2010-056}. 
The detector simulation is based on the $\geant$ package \cite{Agostinelli:2002hh,Allison:2006ve}. 
The BDT classifier uses information on the candidate's kinematic properties, decay length, vertex quality, impact parameter and angle between the particle momentum and the vector from the primary to the secondary vertex. 
The distributions of the variables that are used in the BDT are similar for $\Bcp \to \jpsi \pi^+$ and $\Bcp \to \psitwos \pi^+$ decays. 
The simulated sample of $\Bcp \to \jpsi\pip$ is used as the signal sample for the BDT training.
The main background is combinatorial, and is represented by the upper sideband in the $\Bcp$ mass spectrum
from the $\Bcp \to \jpsi\pip$ data sample,
requiring the reconstructed mass to be in the range $[6346,6444]\mevcc$.
Since the upper sideband is used for the BDT training, the BDT could over-perform in this region and distort the expected combinatorial background in the signal region. 
To avoid possible bias, two BDT classifiers are trained, denoted as \text{BDT1} and \text{BDT2} in the following. 
The $\Bcp \to \jpsi\pip$ simulation and data samples are both split into two halves.
One half of the simulated data sample and of the $\Bcp$ upper sideband is used to train the \text{BDT1} classifier, and the other half for \text{BDT2}.
Each BDT classifier is applied to the other half of the $\Bcp \to \jpsi\pip$ data sample, which is not used for its training. 
The $\Bcp \to \psitwos\pip$ data sample is also split into two sub-samples, one for each BDT classifier.
The threshold value for the BDT response is chosen to maximise the signal significance.
Finally, the $\mup\mun$ invariant mass window $[3030,3170]\mevcc$ is applied to $\jpsi$ candidates, and $[3620,3760]\mevcc$ to $\psitwos$ candidates.

After the full selection, the background in the $\Bcp \to \jpsi\pip$ sample consists of three categories: combinatorial background; partially reconstructed background, mainly from $\Bcp \to \jpsi \rhop$ decays with $\rhop \to \pip\piz$, where the $\piz$ is not reconstructed; and contamination from the Cabibbo-suppressed decay, $\Bcp \to {\jpsi}K^+$, with the kaon misidentified as a pion. 
The background in the $\Bcp \to \psitwos\pip$ sample consists of a combinatorial background and a partially reconstructed background. The contribution from $\Bcp \to \psitwos K^+$ is negligible.

\begin{figure}[t]
\begin{center}
\includegraphics[width=0.48\linewidth]{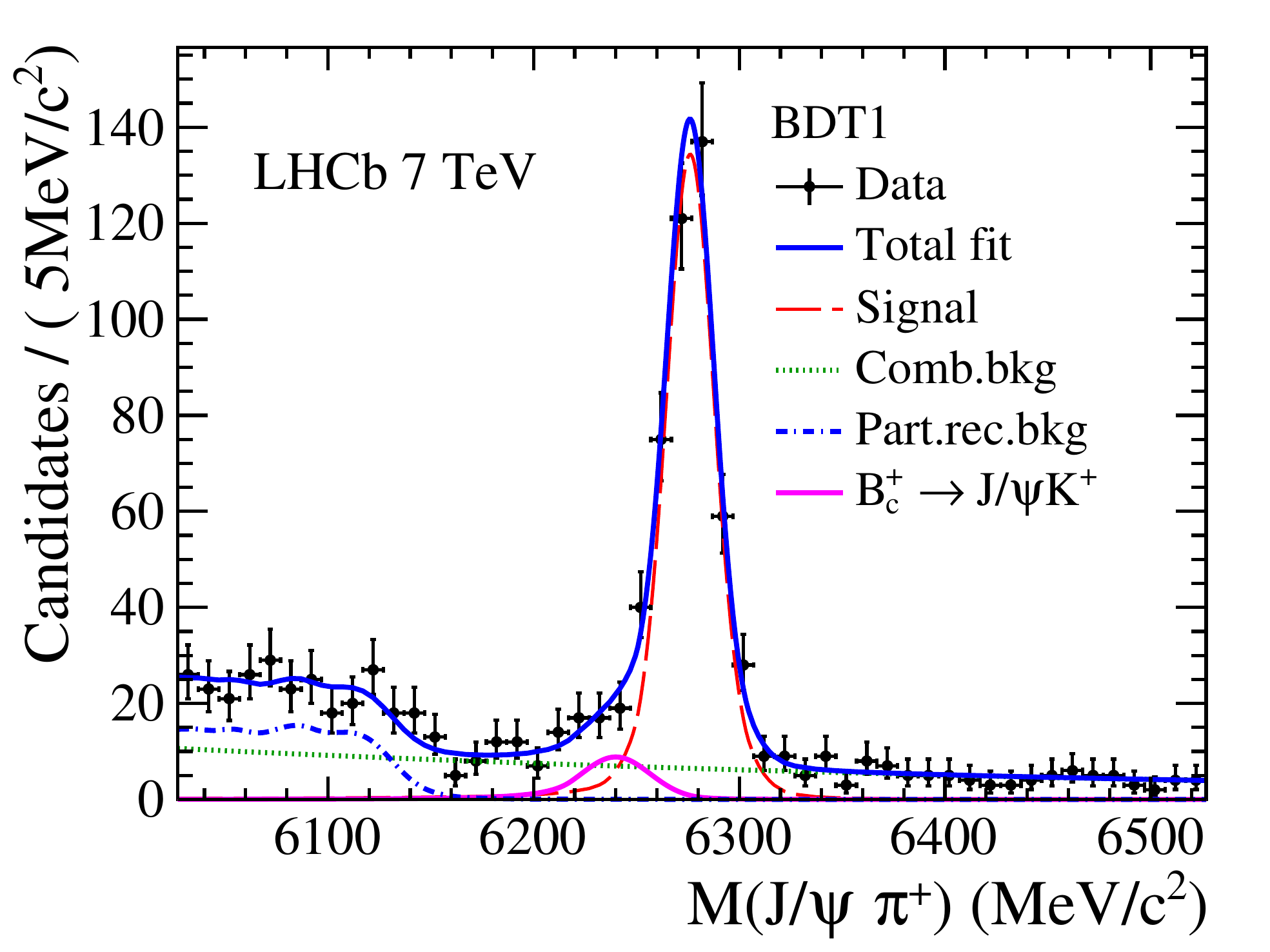}
\includegraphics[width=0.48\linewidth]{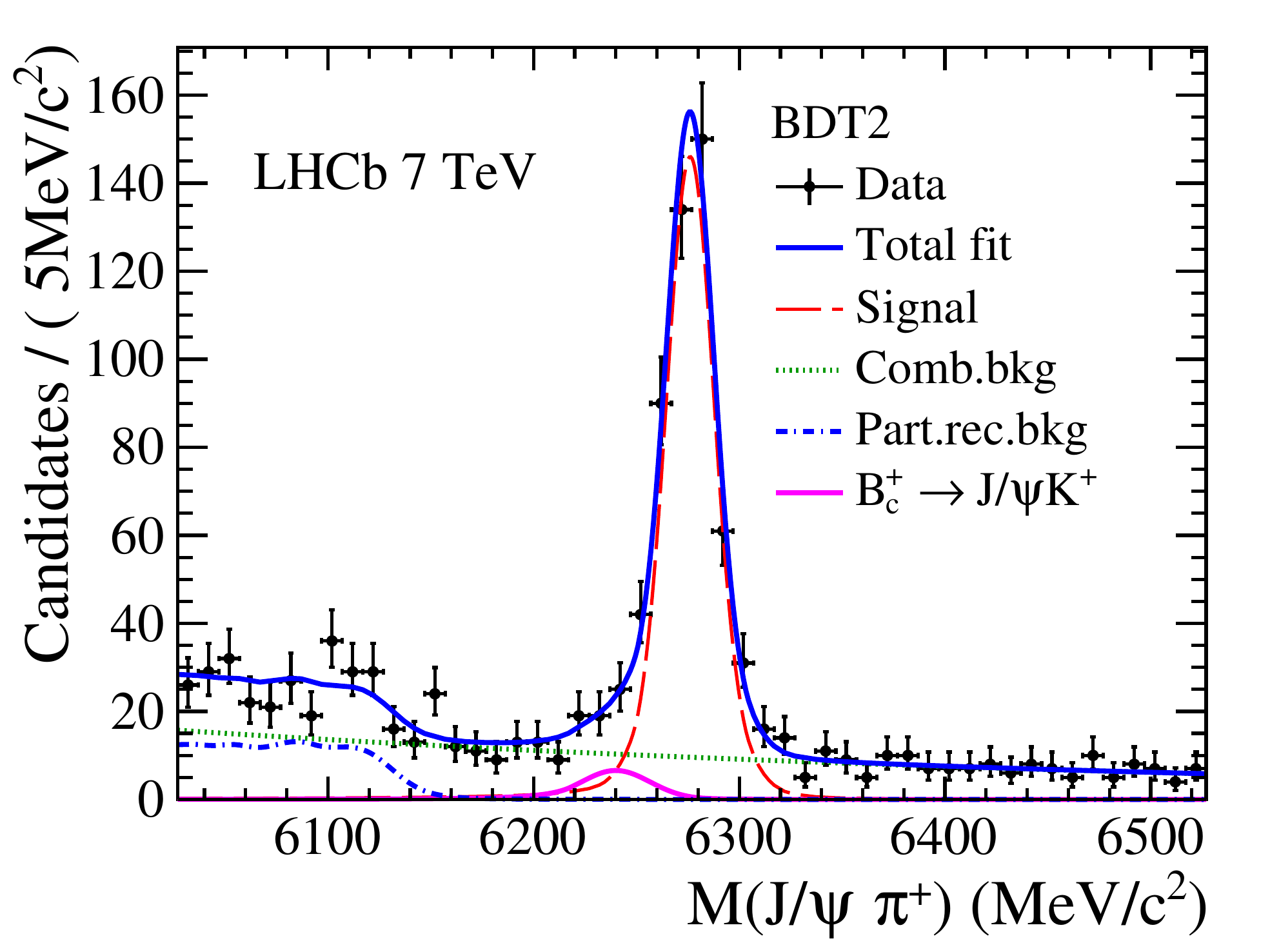}
\includegraphics[width=0.48\linewidth]{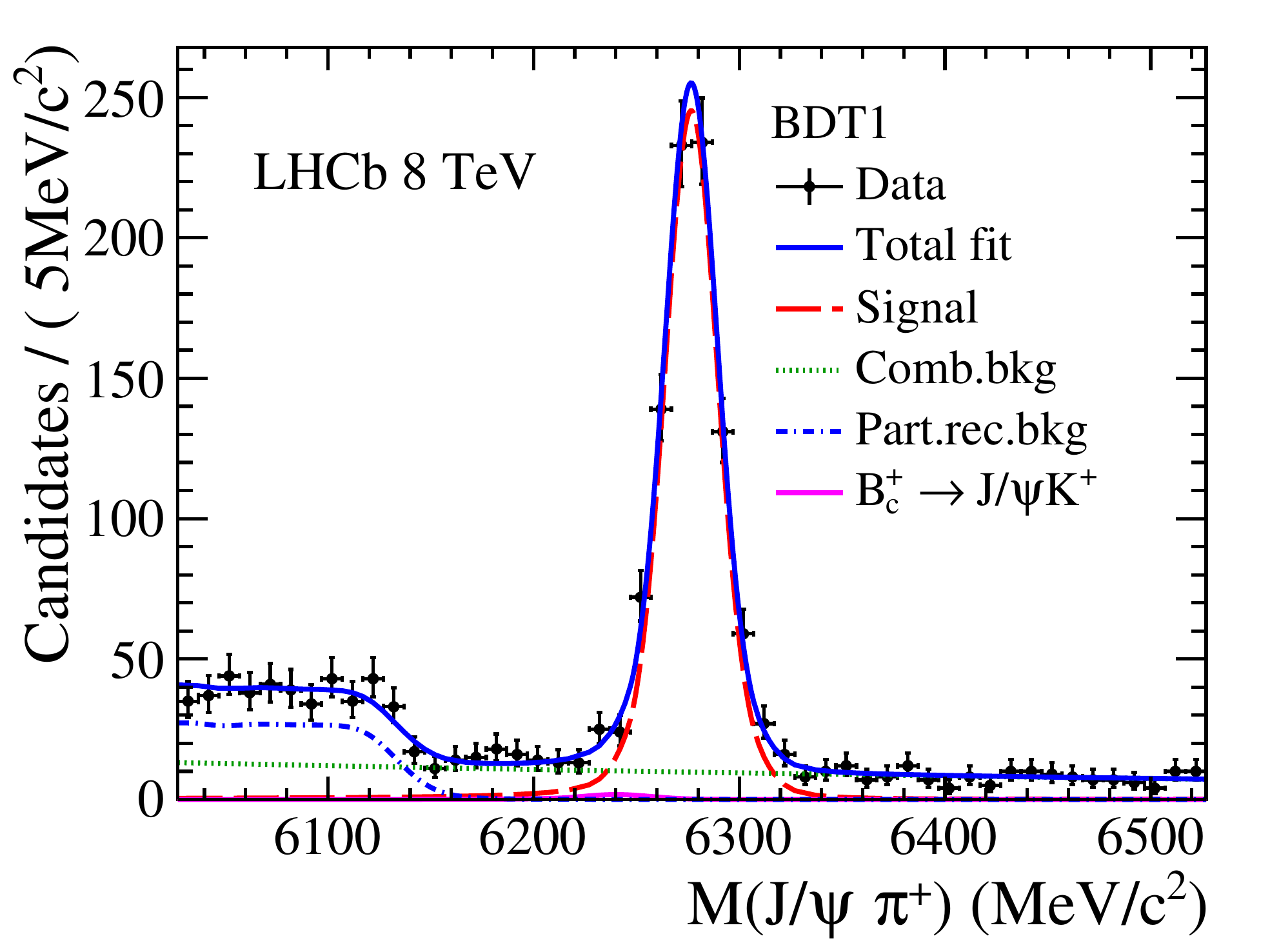}
\includegraphics[width=0.48\linewidth]{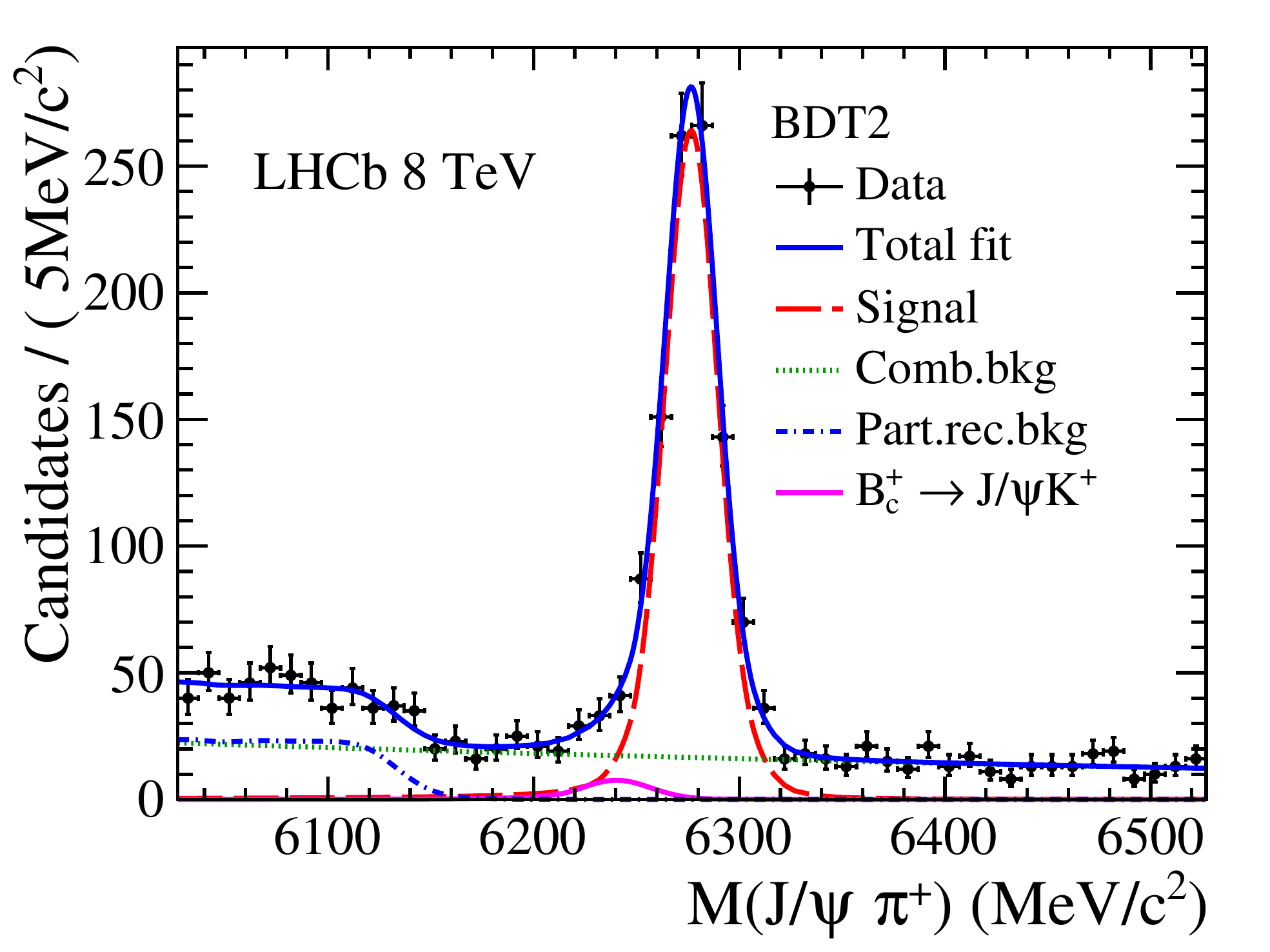}
\vspace*{-0.6cm}
\end{center}
\caption{Fit to the reconstructed $\Bcp$ mass distribution for $\Bcp \to \jpsi\pi^+$ using (top)~2011 and (bottom)~2012 data samples. The plots on the left~(right) correspond to the data selected with \text{BDT1}~(\text{BDT2}). Black points with error bars represent the data, and the various components are indicated in the keys.}\label{fig:fitresultsJ}
\end{figure}

\begin{figure}[t]
\begin{center}
\includegraphics[width=0.48\linewidth]{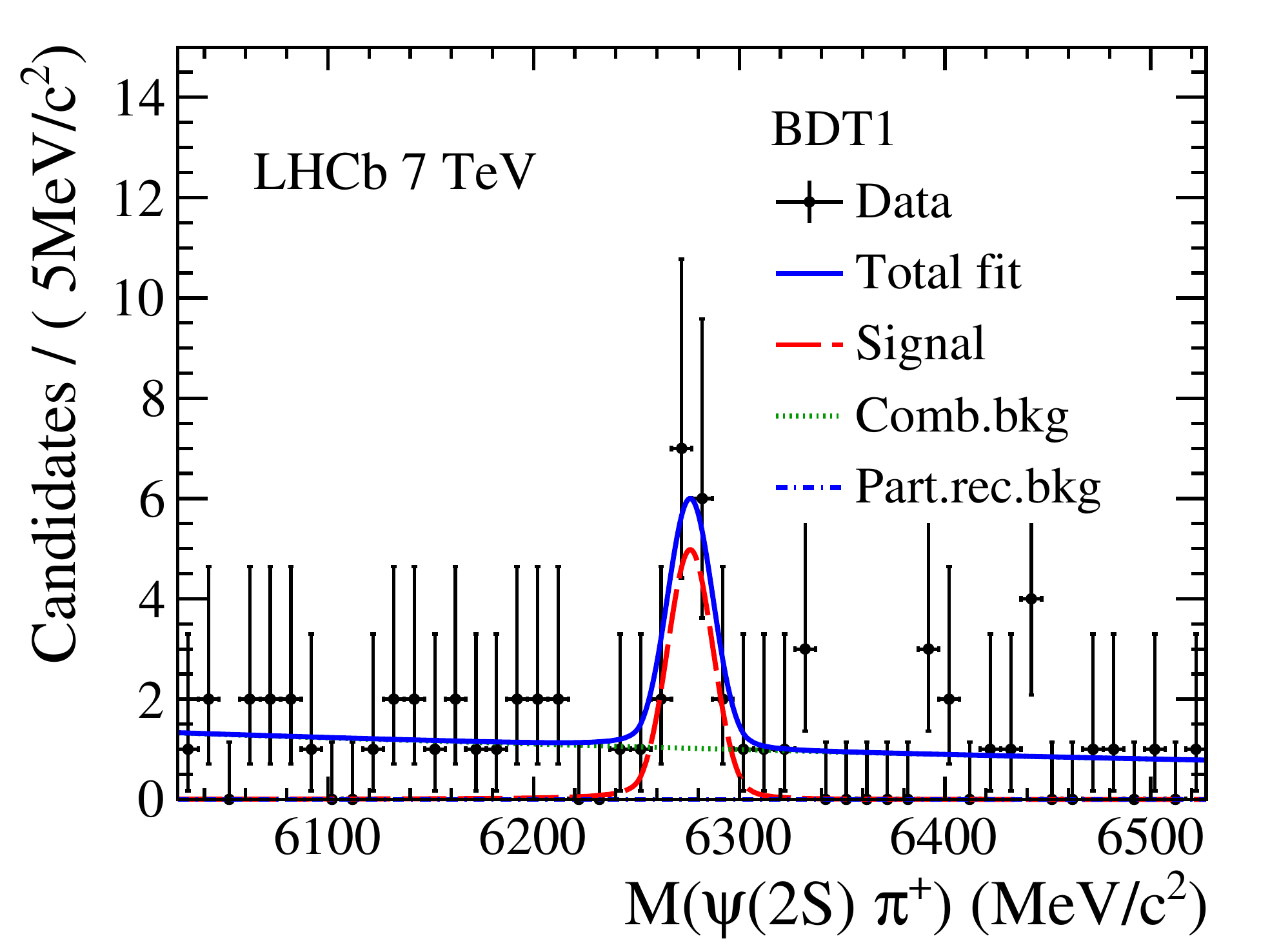}
\includegraphics[width=0.48\linewidth]{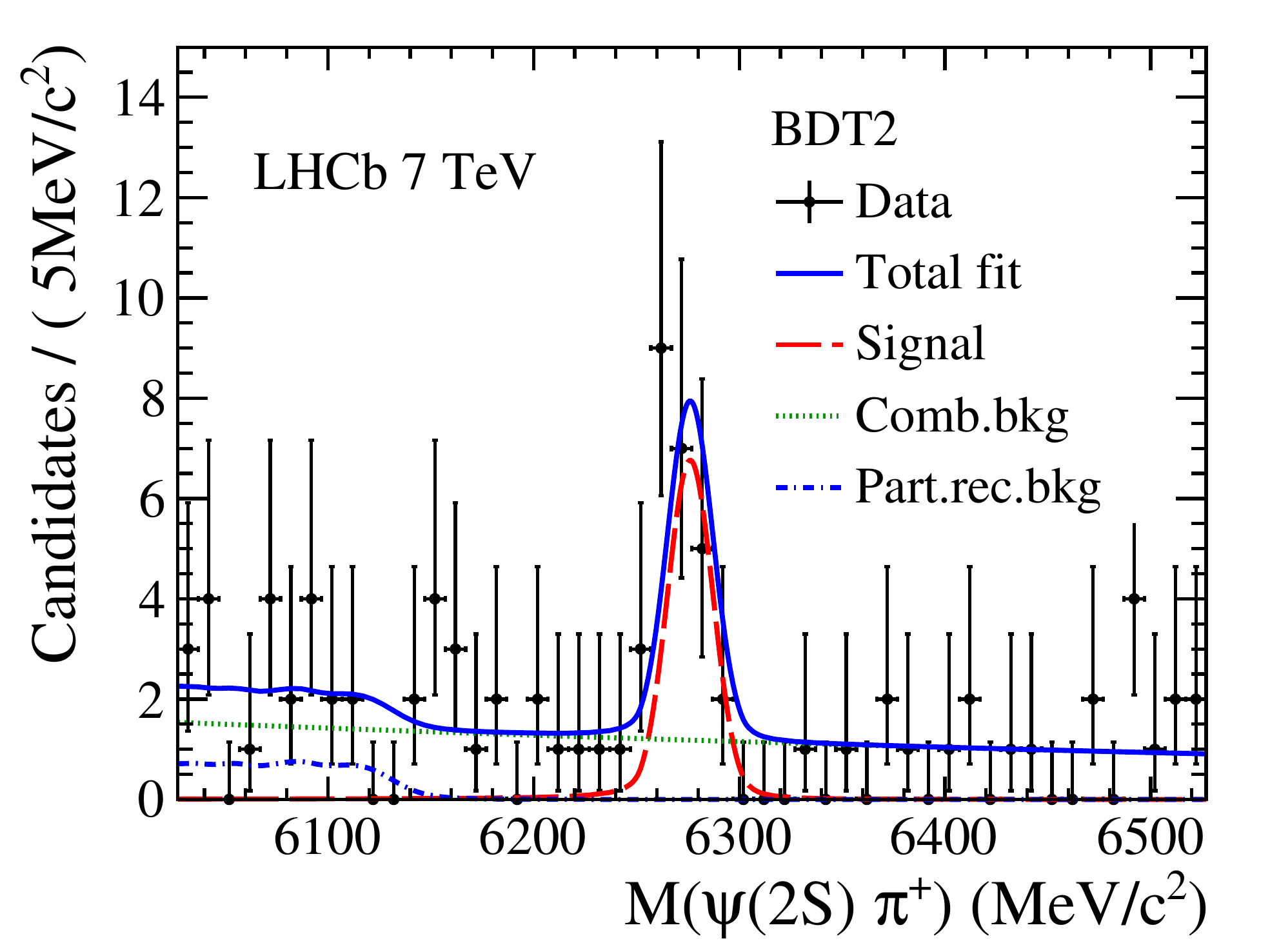}
\includegraphics[width=0.48\linewidth]{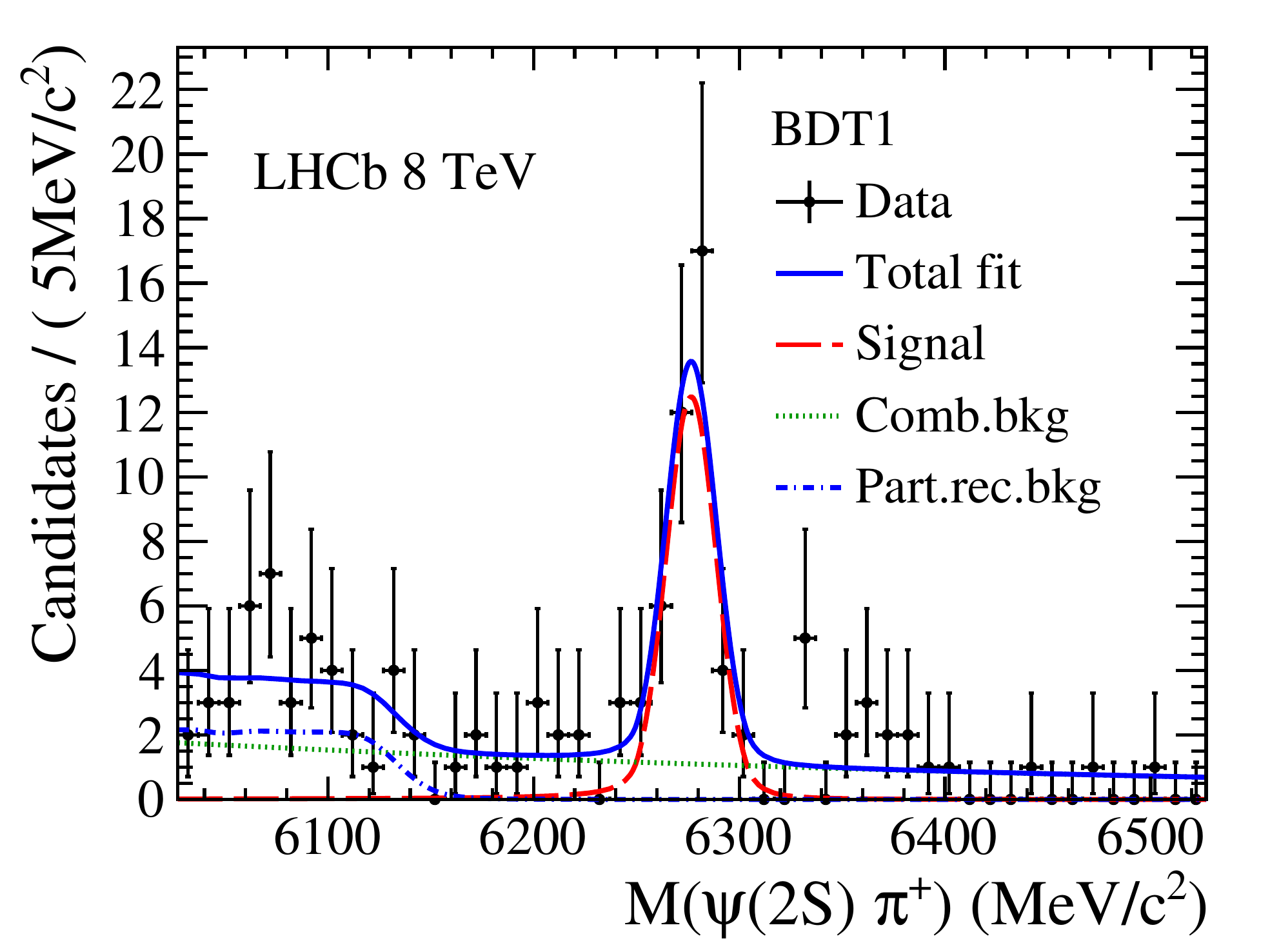}
\includegraphics[width=0.48\linewidth]{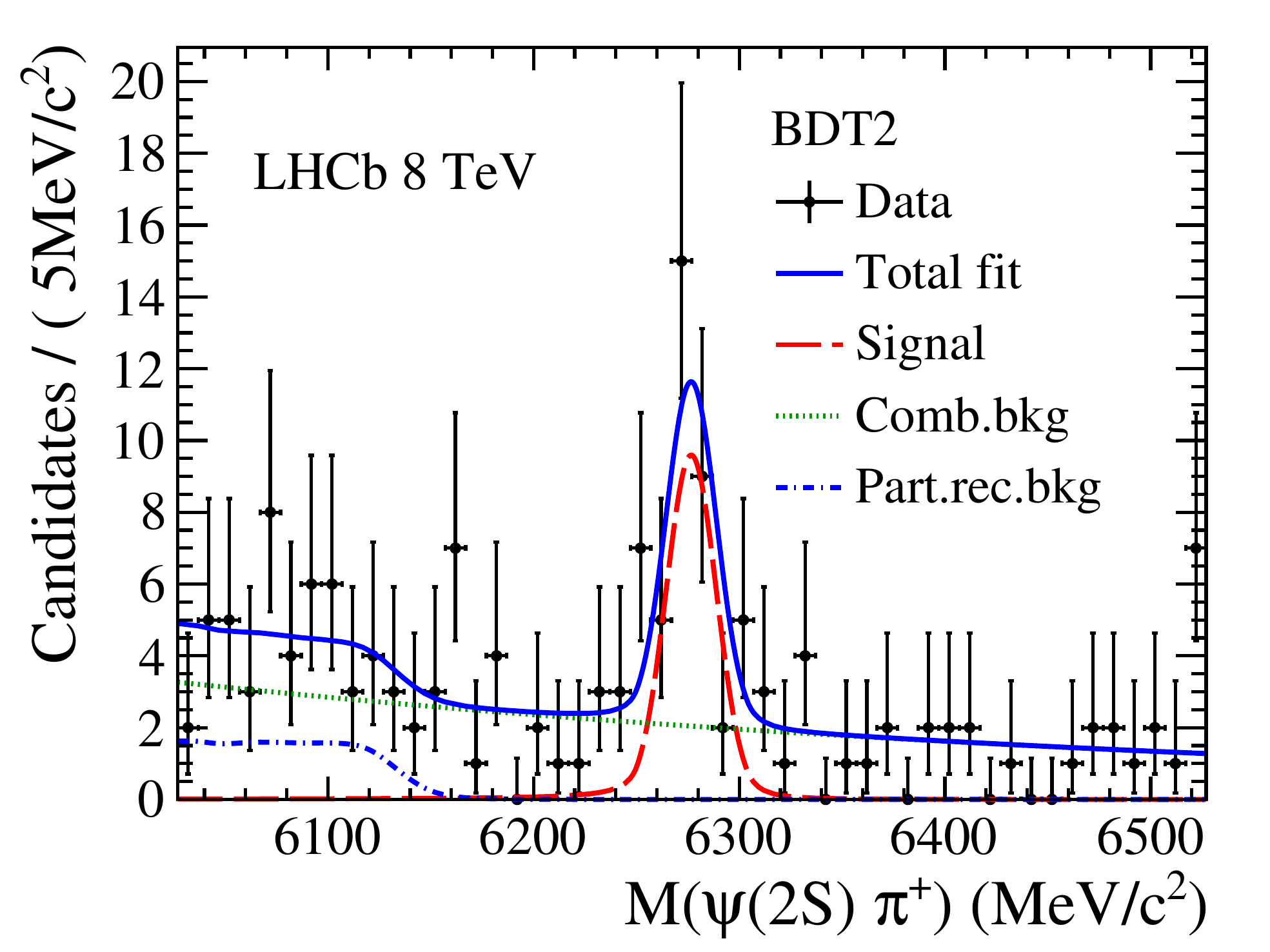}
\vspace*{-0.6cm}
\end{center}
\caption{Fit to the reconstructed $\Bcp$ mass distribution for $\Bcp \to \psitwos\pi^+$ using (top)~2011 and (bottom)~2012 data samples. The plots on the left~(right) correspond to the data selected with \text{BDT1}~(\text{BDT2}). Black points with error bars represent the data, and the various components are indicated in the keys.}\label{fig:fitresultsp}
\end{figure}

The signal yields are extracted from unbinned extended maximum likelihood fits to the invariant mass distributions of $\jpsi\pip$ or $\psitwos\pip$ 
in the range $[6027, 6527] \mevcc$,
as shown in Figs.~\ref{fig:fitresultsJ} and \ref{fig:fitresultsp} for 2011 and 2012 data,
and are summarized in Tables~\ref{tab:tableJ} and~\ref{tab:tablep}. 
To improve the $\Bcp$ mass resolution, the masses of $\jpsi$ and $\psitwos$ candidates are constrained to their known values~\cite{PDG2014}. 
For the $\Bcp \to \jpsi\pip$ channel, the signal probability density function is modelled by the sum of two double-sided Crystal Ball functions~\cite{Skwarnicki:1986xj}, with the same mean value and tail parameters determined from simulation; 
the combinatorial background is described with an exponential function;   
and the partially reconstructed background is modelled with the distribution of the $\Bcp$ invariant mass obtained from a simulated $\Bcp \to \jpsi \rhop$ sample using a kernel estimation~\cite{Cranmer:2000du}.
This last shape is convolved with a Gaussian distribution to take into account a difference in mass resolution between data and simulation.
For the $\Bcp \to {\jpsi}K^+$ background, the shape of the $\Bcp$ mass distribution is modelled by a double-sided Crystal Ball function with parameters determined from simulation.  
For the $\Bcp \to \psitwos \pip$ channel, due to the limited statistics, the signal shape is modelled by a single double-sided Crystal Ball function with the tail parameters determined from simulation; 
the combinatorial and partially reconstructed backgrounds are described with the same models as used for the $\Bcp \to \jpsi\pip$ channel. 
%%%%%%%%%%%%%%%%%%%%%%%%%%%%%%%%%%%%%%%%%%%%%%%%%%
%%%%%%%%%%%%%%%%%%%%%%%%%%%%%%%%%%%%%%%%%%%%%%%%%%%%%%%%%%
%%%%%%%%%%%%%%%%%%%%%%%%%%%%%%%%%%%%%%%%%%%%%%%%%%%%%%%%%%
%%%%%%%%%%%%%%%%%%%%%%%%%%%%%%%%%%%%%%%%%%%
\begin{table}[b]
\caption{Summary of the signal yields and efficiencies for the $\Bcp \to \jpsi\pip$ decay.}\label{tab:tableJ} 
\begin{center}
\renewcommand\arraystretch{1.15}
\vspace*{-0.1cm}
\begin{tabular}{l|cc|cc} 
\hline
 & \multicolumn{2}{c|}{2011} & \multicolumn{2}{c}{2012} \\
 & $\text{BDT1}$ & $\text{BDT2}$ & $\text{BDT1}$ & $\text{BDT2}$ \\
\hline
yield & $437\pm24$ & $475\pm26$ & $883\pm34$ & $950\pm36$ \\ 
$\varepsilon_{\rm{BDT}}$ & $(62.99\pm0.07)\%$ & $(69.29\pm0.06)\%$ & $(62.33\pm0.06)\%$ & $(68.50\pm0.06)\%$ \\ 
$\varepsilon^{*}$ & \multicolumn{2}{c|}{$(1.392\pm0.003)\%$} & \multicolumn{2}{c}{$(1.339\pm0.003)\%$}\\ 
\hline
\end{tabular}
\end{center}
\end{table}
%%%%%%%%%%%%%%%%%%%%%%%%%%%%%%%%%%%%%%%%%%%

The total selection efficiency is the product of the detector geometrical acceptance, the trigger efficiency, the reconstruction and selection efficiency, the PID efficiency, and the BDT classifier efficiency. 
All efficiencies are determined using simulated samples. 
To account for any discrepancy between data and simulation, the PID efficiencies are calibrated using a $\pip$ sample from $D^*$-tagged $D^0 \to K^-\pip$ decays.
The BDT classifier efficiencies of \text{BDT1} and \text{BDT2} are slightly different.
After correcting for the BDT classifier efficiencies the signal yields of the sub-samples are consistent within the statistical uncertainties. 
The BDT classifier efficiency, $\varepsilon_{\rm{BDT}}$, and the product of all other efficiencies, $\varepsilon^{*}$, are listed in Tables~\ref{tab:tableJ} and~\ref{tab:tablep}.

Several sources of systematic uncertainty 
on the $R_{\BF}$ measurement
are studied and are summarized in Table~\ref{tab:sys}.
To account for the uncertainty due to the signal shape modelling, the data are refitted with an alternative shape.   
The $\Bcp$ invariant mass distributions are modelled by a kernel estimation convolved with a Gaussian function, as determined from simulation. 
A difference of $0.6\%$ from the nominal result is observed and is taken as a systematic uncertainty.

The modelling of the partially reconstructed background can also introduce a systematic uncertainty. This is estimated by reducing the fit range to $[6164,6527]\mevcc$ to exclude its contribution. 
A change of $2.4\%$ in the result is observed.
In the nominal fits, the parameters for $\Bcp \to \jpsi K^+$ and the partially reconstructed background are fixed; the results change by less than 1\% when these parameters are allowed to vary. The systematic uncertainty due to background modelling is estimated to be 2.4\%.

Systematic uncertainties on the $\RB$ measurement can be introduced by the BDT classifier efficiency if the simulation fails to describe the data.
The distributions of all training variables from simulation and background-subtracted data are compared, where the background subtraction is performed using the $\sPlot$ technique, taking the $\Bcp$ invariant mass as the discriminating variable~\cite{Pivk:2004ty}.
They are generally in agreement within statistical fluctuations. 
Only one variable, which describes the consistency between the pion track and the PV, indicates small differences between simulation and data. 
Therefore, the simulated sample is reweighed to match the data, and the BDT efficiencies are recalculated with the reweighed simulated sample. The result obtained with these BDT efficiencies is different from the nominal value by $0.2\%$, which is taken as the uncertainty from the BDT classifier.

The efficiences determined from simulated samples have uncertainties due to the limited statistics. This leads to an uncertainty of $0.3\%$.
An uncertainty of $1.1\%$ is assigned due to imperfect simulation of the trigger, which is determined using data driven methods~\cite{LHCb-PUB-2011-016,LHCb-PAPER-2012-010}.
The $\Bcp$ lifetime of simulated samples is set according to the latest LHCb measurement~\cite{LHCb-PAPER-2013-063}. 
To estimate the systematic uncertainty due to this, the $\Bcp$ lifetime is varied within the uncertainty of this measurement, and the change in the result, $0.1\%$, is taken as a systematic uncertainty.
The total systematic uncertainty is $2.7\%$.

The ratio of the branching fractions with $\jpsi$ and $\psitwos$ mesons decaying to dimuons, denoted as 
\begin{equation}
R\equiv\frac{\BR(\Bcp \to \psitwos\pi^+, \psitwos \to \mu^+\mu^-)}{\BR(\Bcp \to \jpsi\pi^+, \jpsi \to \mu^+\mu^-)},
\end{equation}
is calculated as 
\begin{equation}
R = \frac{N_{2011}^{\rm{cor}}(\Bcp \to \psitwos\pip)+N_{2012}^{\rm{cor}}(\Bcp \to \psitwos\pip)}{N_{2011}^{\rm{cor}}(\Bcp \to \jpsi\pip)+N_{2012}^{\rm{cor}}(\Bcp \to \jpsi\pip)},
\end{equation}
where $N_{2011(2012)}^{\rm{cor}}$ are the signal yields from 2011(2012) after efficiency correction.
The ratio is measured to be 
\begin{equation*}
R = 0.0354\pm0.0042\stat\pm0.0010\syst.
\end{equation*}
The ratio of the branching fractions of $\Bcp \to \psitwos\pi^+$ and $\Bcp \to \jpsi\pi^+$ is calculated as
\begin{equation}
\RB = R \times \frac{\BR(\jpsi \to \mu^+\mu^-)}{\BR(\psitwos \to \mu^+\mu^-)}.
\end{equation}
%%%%%%%%%%%%%%%%%%%%%%%%%%%%%%%%%%%%%%%
\begin{table}[t]
\caption{Summary of the signal yields and efficiencies for the $\Bcp \to \psitwos\pip$ decay.}\label{tab:tablep} 
\begin{center}
\renewcommand\arraystretch{1.15}
\vspace*{-0.1cm}
\begin{tabular}{l|cc|cc} 
\hline
 & \multicolumn{2}{c|}{2011} & \multicolumn{2}{c}{2012} \\
 & $\text{BDT1}$ & $\text{BDT2}$ & $\text{BDT1}$ & $\text{BDT2}$ \\
\hline
yield & $14.4\pm4.5$ & $19.6\pm5.3$ & $40.1\pm7.1$ & $30.8\pm7.0$ \\ 
$\varepsilon_{\rm{BDT}}$ & $(58.79\pm0.11)\%$ & $(65.84\pm0.11)\%$ & $(58.32\pm0.08)\%$ & $(65.08\pm0.08)\%$ \\ 
$\varepsilon^{*}$ & \multicolumn{2}{c|}{$(1.631\pm0.006)\%$} & \multicolumn{2}{c}{$(1.529\pm0.005)\%$}\\ 
\hline
\end{tabular}
\end{center}
\end{table}
%%%%%%%%%%%%%%%%%%%%%%%%%%%%%%%%%%%%%%%
\begin{table}[t]
\caption{Summary of systematic uncertainties on the $\RB$ measurement.}
\begin{center}
\vspace*{-0.1cm}
\begin{tabular}{lc}
\hline
Component & Uncertainty \\
\hline
Signal shape & $0.6\%$ \\
Background shape & $2.4\%$ \\
BDT classifier & $0.2\%$ \\
Monte-Carlo Statistics & $0.3\%$ \\
Trigger efficiency & $1.1\%$ \\
$\Bcp$ lifetime & $0.1\%$ \\
\hline
Total & $2.7\%$ \\
\hline
\end{tabular}\end{center}
\label{tab:sys}
\end{table}
%%%%%%%%%%%%%%%%%%%%%%%%%%%%%%%%%%%%%%%

Assuming electroweak universality, the $\jpsi \to \mup\mun$ and $\psitwos \to \mup\mun$ branching fractions can be substituted with the more precisely measured ones in the $e^+e^-$ channel~\cite{PDG2014}.
Using these values, the ratio $\RB$ is measured to be: 
\begin{equation*}
\RB = \NewRes,
\end{equation*}
\noindent where the first uncertainty is statistical, the second is systematic, and the last term is due to the uncertainty on $\BR(\jpsi \to e^+e^-)/\BR(\psitwos \to e^+e^-)$.
This result is in agreement with the previous $\lhcb$ result~\cite{LHCb-PAPER-2012-054}.
Our measurement is consistent with the predictions of non-relativistic QCD at next-to-leading order ($0.26^{+0.05}_{-0.06}$)~\cite{Qiao:2012hp} and perturbative QCD based on $k_{\rm{T}}$ factorization ($0.29^{+0.17}_{-0.11}$)~\cite{Rui:2015iia}.
The result disfavors the theoretical calculations based on the relativistic quark model~\cite{Ebert:2003cn}, the quark potential model~\cite{Chang:1992pt}, the relativistic constituent quark model~\cite{Liu:1997hr} and the QCD relativistic potential model~\cite{Colangelo:1999zn}.

\section*{Acknowledgements}
\noindent We express our gratitude to our colleagues in the CERN
accelerator departments for the excellent performance of the LHC. We
thank the technical and administrative staff at the LHCb
institutes. We acknowledge support from CERN and from the national
agencies: CAPES, CNPq, FAPERJ and FINEP (Brazil); NSFC (China);
CNRS/IN2P3 (France); BMBF, DFG, HGF and MPG (Germany); INFN (Italy); 
FOM and NWO (The Netherlands); MNiSW and NCN (Poland); MEN/IFA (Romania); 
MinES and FANO (Russia); MinECo (Spain); SNSF and SER (Switzerland); 
NASU (Ukraine); STFC (United Kingdom); NSF (USA).
The Tier1 computing centres are supported by IN2P3 (France), KIT and BMBF 
(Germany), INFN (Italy), NWO and SURF (The Netherlands), PIC (Spain), GridPP 
(United Kingdom).
We are indebted to the communities behind the multiple open 
source software packages on which we depend. We are also thankful for the 
computing resources and the access to software R\&D tools provided by Yandex LLC (Russia).
Individual groups or members have received support from 
EPLANET, Marie Sk\l{}odowska-Curie Actions and ERC (European Union), 
Conseil g\'{e}n\'{e}ral de Haute-Savoie, Labex ENIGMASS and OCEVU, 
R\'{e}gion Auvergne (France), RFBR (Russia), XuntaGal and GENCAT (Spain), Royal Society and Royal
Commission for the Exhibition of 1851 (United Kingdom).

\addcontentsline{toc}{section}{References}
\setboolean{inbibliography}{true}
%\bibliographystyle{LHCb}
%\bibliography{main,LHCb-PAPER,LHCb-CONF,LHCb-DP,LHCb-TDR}

%%%%%%%%%%%%%%%%%%%%%%%%%
%%%%% Bibliography %%%%%%
%%%%%%%%%%%%%%%%%%%%%%%%%
\ifx\mcitethebibliography\mciteundefinedmacro
\PackageError{LHCb.bst}{mciteplus.sty has not been loaded}
{This bibstyle requires the use of the mciteplus package.}\fi
\providecommand{\href}[2]{#2}

\newpage

% Author List ----------------------------                                                                                                                                                                                                                                                                                                
%  You need to get a new author list!                                                                                                                                                                                                                                                                                                    

%\input{LHCb_HD_authorlist_2014-06-20}
 
\newpage
%\input{LHCb_authorlist.tex}
\input{LHCb_HD_authorlist_2015-05-11}

\end{document}

%% file: preamble.tex
\usepackage{microtype}
\usepackage{lineno}  % for line numbering during review
\usepackage{xspace} % To avoid problems with missing or double spaces after
\usepackage{caption} %these three command get the figure and table captions automatically small

\usepackage{graphicx}  % to include figures (can also use other packages)
\usepackage{color}
\usepackage{colortbl}
\graphicspath{{./figs/}} % Make Latex search fig subdir for figures

\usepackage{amsmath} % Adds a large collection of math symbols
\usepackage{amssymb}
\usepackage{amsfonts}
\usepackage{upgreek} % Adds in support for greek letters in roman typeset

\newcommand*\patchAmsMathEnvironmentForLineno[1]{%
\expandafter\let\csname old#1\expandafter\endcsname\csname #1\endcsname
\expandafter\let\csname oldend#1\expandafter\endcsname\csname
end#1\endcsname
 \renewenvironment{#1}%
   {\linenomath\csname old#1\endcsname}%
   {\csname oldend#1\endcsname\endlinenomath}%
}
\newcommand*\patchBothAmsMathEnvironmentsForLineno[1]{%
  \patchAmsMathEnvironmentForLineno{#1}%
  \patchAmsMathEnvironmentForLineno{#1*}%
}
\AtBeginDocument{%
\patchBothAmsMathEnvironmentsForLineno{equation}%
\patchBothAmsMathEnvironmentsForLineno{align}%
\patchBothAmsMathEnvironmentsForLineno{flalign}%
\patchBothAmsMathEnvironmentsForLineno{alignat}%
\patchBothAmsMathEnvironmentsForLineno{gather}%
\patchBothAmsMathEnvironmentsForLineno{multline}%
\patchBothAmsMathEnvironmentsForLineno{eqnarray}%
}

\usepackage{hyperref}    % Hyperlinks in references
\usepackage[all]{hypcap} % Internal hyperlinks to floats.

\input{lhcb-symbols-def} % Add in the predefined LHCb symbols

\usepackage{cite} % Allows for ranges in citations
\usepackage{mciteplus}

%% file: lhcb-symbols-def.tex
\def\lhcb {\mbox{LHCb}\xspace}

\def\MagUp {\mbox{\em Mag\kern -0.05em Up}\xspace}

\ifthenelse{\boolean{uprightparticles}}%
{

 \def\Pmu         {\ensuremath{\upmu}\xspace}

 \def\Ppi         {\ensuremath{\uppi}\xspace}                 
                  
 \def\Prho        {\ensuremath{\uprho}\xspace}

 \def\Ppsi        {\ensuremath{\uppsi}\xspace}

 \def\PDelta      {\ensuremath{\Delta}\xspace}                 
 \def\PXi      {\ensuremath{\Xi}\xspace}                 
 \def\PLambda      {\ensuremath{\Lambda}\xspace}                 
 \def\PSigma      {\ensuremath{\Sigma}\xspace}                 
 \def\POmega      {\ensuremath{\Omega}\xspace}                 
 \def\PUpsilon      {\ensuremath{\Upsilon}\xspace}                 
 
 \def\PB      {\ensuremath{\mathrm{B}}\xspace}                 
                  
 \def\PD      {\ensuremath{\mathrm{D}}\xspace}

 \def\PJ      {\ensuremath{\mathrm{J}}\xspace}                 
 \def\PK      {\ensuremath{\mathrm{K}}\xspace}

 \def\PW      {\ensuremath{\mathrm{W}}\xspace}

 \def\Pb      {\ensuremath{\mathrm{b}}\xspace}                 
 \def\Pc      {\ensuremath{\mathrm{c}}\xspace}

 \def\Pi      {\ensuremath{\mathrm{i}}\xspace}

}
{

 \def\Pmu         {\ensuremath{\mu}\xspace}

 \def\Ppi         {\ensuremath{\pi}\xspace}                 
                  
 \def\Prho        {\ensuremath{\rho}\xspace}

 \def\Ppsi        {\ensuremath{\psi}\xspace}                 
                  
 \mathchardef\PDelta="7101
 \mathchardef\PXi="7104
 \mathchardef\PLambda="7103
 \mathchardef\PSigma="7106
 \mathchardef\POmega="710A
 \mathchardef\PUpsilon="7107
                  
 \def\PB      {\ensuremath{B}\xspace}                 
                  
 \def\PD      {\ensuremath{D}\xspace}

 \def\PJ      {\ensuremath{J}\xspace}                 
 \def\PK      {\ensuremath{K}\xspace}

 \def\PW      {\ensuremath{W}\xspace}

 \def\Pb      {\ensuremath{b}\xspace}                 
 \def\Pc      {\ensuremath{c}\xspace}

 \def\Pi      {\ensuremath{i}\xspace}

}

\makeatletter
\ifcase \@ptsize \relax% 10pt
  \newcommand{\miniscule}{\@setfontsize\miniscule{4}{5}}% \tiny: 5/6
\or% 11pt
  \newcommand{\miniscule}{\@setfontsize\miniscule{5}{6}}% \tiny: 6/7
\or% 12pt
  \newcommand{\miniscule}{\@setfontsize\miniscule{5}{6}}% \tiny: 6/7
\fi
\makeatother

\DeclareRobustCommand{\optbar}[1]{\shortstack{{\miniscule (\rule[.5ex]{1.25em}{.18mm})}
  \\ [-.7ex] $#1$}}

\def\mup        {{\ensuremath{\Pmu^+}}\xspace}
\def\mun        {{\ensuremath{\Pmu^-}}\xspace} % muon negative (\mum is taken)
\def\mumu       {{\ensuremath{\Pmu^+\Pmu^-}}\xspace}

\def\W      {{\ensuremath{\PW}}\xspace}

\def\cquark    {{\ensuremath{\Pc}}\xspace}
\def\cquarkbar {{\ensuremath{\overline \cquark}}\xspace}

\def\bquark    {{\ensuremath{\Pb}}\xspace}

\def\pion   {{\ensuremath{\Ppi}}\xspace}
\def\piz    {{\ensuremath{\pion^0}}\xspace}

\def\pip    {{\ensuremath{\pion^+}}\xspace}

\def\rhomeson {{\ensuremath{\Prho}}\xspace}

\def\rhop     {{\ensuremath{\rhomeson^+}}\xspace}

  \def\Kbar    {{\kern 0.2em\overline{\kern -0.2em \PK}{}}\xspace}

\def\KorKbar    {\kern 0.18em\optbar{\kern -0.18em K}{}\xspace}

  \def\Dbar    {{\kern 0.2em\overline{\kern -0.2em \PD}{}}\xspace}

\def\DorDbar    {\kern 0.18em\optbar{\kern -0.18em D}{}\xspace}

\def\B       {{\ensuremath{\PB}}\xspace}
\def\Bbar    {{\ensuremath{\kern 0.18em\overline{\kern -0.18em \PB}{}}}\xspace}

\def\BorBbar    {\kern 0.18em\optbar{\kern -0.18em B}{}\xspace}

\def\Bcp     {{\ensuremath{\B_\cquark^+}}\xspace}

\def\jpsi     {{\ensuremath{{\PJ\mskip -3mu/\mskip -2mu\Ppsi\mskip 2mu}}}\xspace}
\def\psitwos  {{\ensuremath{\Ppsi{(2S)}}}\xspace}

  \def\Y#1S{\ensuremath{\PUpsilon{(#1S)}}\xspace}% no space before {...}!

\def\Lbar        {{\ensuremath{\kern 0.1em\overline{\kern -0.1em\PLambda}}}\xspace}
\def\LorLbar    {\kern 0.18em\optbar{\kern -0.18em \PLambda}{}\xspace}

\def\BF         {{\ensuremath{\cal B}}\xspace}

\def\BR         {\BF}
\newcommand{\decay}[2]{\ensuremath{#1\!\to #2}\xspace}         % {\Pa}{\Pb \Pc}

\def\to                 {\ensuremath{\rightarrow}\xspace}

\def\AT#1     {\ensuremath{A_{\mathrm{T}}^{#1}}\xspace}           % 2

\def\C#1      {\ensuremath{\mathcal{C}_{#1}}\xspace}                       % 9
\def\Cp#1     {\ensuremath{\mathcal{C}_{#1}^{'}}\xspace}                    % 7
\def\Ceff#1   {\ensuremath{\mathcal{C}_{#1}^{\mathrm{(eff)}}}\xspace}        % 9  
\def\Cpeff#1  {\ensuremath{\mathcal{C}_{#1}^{'\mathrm{(eff)}}}\xspace}       % 7
\def\Ope#1    {\ensuremath{\mathcal{O}_{#1}}\xspace}                       % 2
\def\Opep#1   {\ensuremath{\mathcal{O}_{#1}^{'}}\xspace}                    % 7

\newcommand{\tev}{\ifthenelse{\boolean{inbibliography}}{\ensuremath{~T\kern -0.05em eV}\xspace}{\ensuremath{\mathrm{\,Te\kern -0.1em V}}}\xspace}
\newcommand{\gev}{\ensuremath{\mathrm{\,Ge\kern -0.1em V}}\xspace}
\newcommand{\mev}{\ensuremath{\mathrm{\,Me\kern -0.1em V}}\xspace}
\newcommand{\kev}{\ensuremath{\mathrm{\,ke\kern -0.1em V}}\xspace}
\newcommand{\ev}{\ensuremath{\mathrm{\,e\kern -0.1em V}}\xspace}
\newcommand{\gevc}{\ensuremath{{\mathrm{\,Ge\kern -0.1em V\!/}c}}\xspace}
\newcommand{\mevc}{\ensuremath{{\mathrm{\,Me\kern -0.1em V\!/}c}}\xspace}
\newcommand{\gevcc}{\ensuremath{{\mathrm{\,Ge\kern -0.1em V\!/}c^2}}\xspace}
\newcommand{\gevgevcccc}{\ensuremath{{\mathrm{\,Ge\kern -0.1em V^2\!/}c^4}}\xspace}
\newcommand{\mevcc}{\ensuremath{{\mathrm{\,Me\kern -0.1em V\!/}c^2}}\xspace}

\def\mum  {\ensuremath{{\,\upmu\rm m}}\xspace}

\def\invfb   {\ensuremath{\mbox{\,fb}^{-1}}\xspace}

\newcommand{\stat}{\ensuremath{\mathrm{\,(stat)}}\xspace}
\newcommand{\syst}{\ensuremath{\mathrm{\,(syst)}}\xspace}

\def\gsim{{~\raise.15em\hbox{$>$}\kern-.85em
          \lower.35em\hbox{$\sim$}~}\xspace}
\def\lsim{{~\raise.15em\hbox{$<$}\kern-.85em
          \lower.35em\hbox{$\sim$}~}\xspace}

\def\sPlot{\mbox{\em sPlot}\xspace}
\def\sqs   {\ensuremath{\protect\sqrt{s}}\xspace}

\def\ptot       {\mbox{$p$}\xspace}
\def\pt         {\mbox{$p_{\rm T}$}\xspace}

\def\bcvegpy    {\mbox{\textsc{Bcvegpy}}\xspace}

\def\geant      {\mbox{\textsc{Geant4}}\xspace}

\def\pythia     {\mbox{\textsc{Pythia}}\xspace}

\def\tell1  {TELL1\xspace}
\def\ukl1   {UKL1\xspace}

%% file: mydefine.tex
\newcommand{\xx}{\ensuremath{\kern 0.5em }}

\def\BcToJpsiPi     {\decay{\Bcp}{\jpsi\pip}}
\def\BcToPsiPi      {\decay{\Bcp}{\psitwos\pip}}
\def\JpsiToMuMu  {\decay{\jpsi}{\mumu}}
\def\PsiToMuMu  {\decay{\psitwos}{\mumu}}

\def\RB         {{\ensuremath{R_{\BF}}\xspace}}
\newcommand{\NewRes}{\ensuremath{0.268 \pm 0.032\stat \pm 0.007\syst \pm 0.006\,(\mathrm{BF}) }}
\newcommand{\OldRes}{\ensuremath{0.250 \pm 0.068\stat \pm 0.014\syst \pm 0.006\,(\mathrm{BF}) }}

\def\sPlot{\mbox{\em sPlot}\xspace}

\usepackage{multirow} % for complicated table
\usepackage{booktabs} % for complicated table
\usepackage{rotating}

%% file: title-LHCb-PAPER.tex
% $Id: title-LHCb-PAPER.tex 76509 2015-07-10 07:56:55Z yiming $
% ===============================================================================
% Purpose: LHCb-PAPER journal paper title page template
% Author: 
% Created on: 2010-09-25
% ===============================================================================

%%%%%%%%%%%%%%%%%%%%%%%%%
%%%%%  TITLE PAGE  %%%%%%
%%%%%%%%%%%%%%%%%%%%%%%%%
\begin{titlepage}
\pagenumbering{roman}

% Header ---------------------------------------------------
\vspace*{-1.5cm}
\centerline{\large EUROPEAN ORGANIZATION FOR NUCLEAR RESEARCH (CERN)}
\vspace*{1.5cm}
\hspace*{-0.5cm}
\begin{tabular*}{\linewidth}{lc@{\extracolsep{\fill}}r}
%\ifthenelse{\boolean{pdflatex}}% Logo format choice
%{\vspace*{-2.7cm}\mbox{\!\!\!\includegraphics[width=.14\textwidth]{lhcb-logo.pdf}} & &}%
%{\vspace*{-1.2cm}\mbox{\!\!\!\includegraphics[width=.12\textwidth]{lhcb-logo.eps}} & &}%
%\\
 & & CERN-PH-EP-2015-167 \\  % ID 
 & & LHCb-PAPER-2015-024 \\  % ID 
 & & 21 October 2015
% & & 13 July 2015 \\ % Date - Can also hardwire e.g.: 23 March 2010
% & & version 3.2\\
% not in paper \hline
\end{tabular*}

\vspace*{3.0cm}

% Title --------------------------------------------------
{\bf\boldmath\LARGE
\begin{center}
  Measurement of the branching fraction ratio \\
  $\BF(\Bcp \to \psitwos \pip)/\BF(\Bcp \to \jpsi\pip)$
\end{center}
}

\vspace*{2.0cm}

% Authors -------------------------------------------------
\begin{center}
%In the footnote, replace 'paper' by 'letter' in case of submission to PRL or PLB 
The LHCb collaboration\footnote{Authors are listed at the end of this paper.}
\end{center}

\vspace{\fill}

% Abstract -----------------------------------------------
\begin{abstract}
  \noindent
Using $pp$ collision data collected by $\lhcb$ at center-of-mass energies $\sqs=7\tev$ and $8\tev$, 
corresponding to an integrated luminosity of $3\invfb$, 
the ratio of the branching fraction of the $\BcToPsiPi$ decay 
relative to that of the $\BcToJpsiPi$ decay is measured to be $\NewRes$.
The first uncertainty is statistical, the second is systematic, 
and the third is due to the uncertainties on the branching fractions of the $\JpsiToMuMu$ and $\PsiToMuMu$ decays.
This measurement is consistent with the previous $\lhcb$ result,
and the statistical uncertainty is halved.

\end{abstract}

\vspace*{2.0cm}

\begin{center}
  Published in Phys.~Rev.~D  
\end{center}

\vspace{\fill}

{\footnotesize 
\centerline{\copyright~CERN on behalf of the \lhcb collaboration, licence \href{http://creativecommons.org/licenses/by/4.0/}{CC-BY-4.0}.}}
\vspace*{2mm}

\end{titlepage}

%%%%%%%%%%%%%%%%%%%%%%%%%%%%%%%%
%%%%%  EOD OF TITLE PAGE  %%%%%%
%%%%%%%%%%%%%%%%%%%%%%%%%%%%%%%%

%  empty page follows the title page ----
\newpage
\setcounter{page}{2}
\mbox{~}
%\newpage
%
%% Author List ----------------------------
%%  You need to get a new author list!
%\input{LHCb_authorlist.tex}
%
%The author list for journal publications is provided by the Membership Committee shortly after 'approval to go to paper' has been given.
%%It will be made available on the page
%%\verb!http://www.physik.uzh.ch/~strauman/forMemCo/LHCb-PAPER-XXXX-XXX/! .
%It will be sent to you by email shortly after a paper number has beens assigned.
%The author list should be included already at first circulation, 
%to allow new members of the collaboration to verify whether they have been included correctly.
%Occasionally a misspelled name is corrected or associated institutions become full members.
%In that case, a new author list will be sent to you.
%In case line numbering doesn't work well after including the authorlist, try moving the \verb!\bigskip! after the last author to a separate line.
%
%
%The authorship for Conference Reports should be ``The LHCb
%  collaboration'', with a footnote giving the name(s) of the contact
%  author(s), but without the full list of collaboration names.

\cleardoublepage

%% file: LHCb_HD_authorlist_2015-05-11.tex
%%%%%%%%%%%%%%%%%%%%%%%%%%%%%%%%%%%%%%%%%%
\centerline{\large\bf LHCb collaboration}
\begin{flushleft}
\small
R.~Aaij$^{38}$, 
B.~Adeva$^{37}$, 
M.~Adinolfi$^{46}$, 
A.~Affolder$^{52}$, 
Z.~Ajaltouni$^{5}$, 
S.~Akar$^{6}$, 
J.~Albrecht$^{9}$, 
F.~Alessio$^{38}$, 
M.~Alexander$^{51}$, 
S.~Ali$^{41}$
% ...

{\footnotesize \it
% ...
$ ^{5}$Clermont Universit\'{e}, Universit\'{e} Blaise Pascal, CNRS/IN2P3, LPC, Clermont-Ferrand, France\\
$ ^{6}$CPPM, Aix-Marseille Universit\'{e}, CNRS/IN2P3, Marseille, France\\
% ...
$ ^{9}$Fakult\"{a}t Physik, Technische Universit\"{a}t Dortmund, Dortmund, Germany\\
% ...
$ ^{37}$Universidad de Santiago de Compostela, Santiago de Compostela, Spain\\
$ ^{38}$European Organization for Nuclear Research (CERN), Geneva, Switzerland\\
% ...
$ ^{41}$Nikhef National Institute for Subatomic Physics, Amsterdam, The Netherlands\\
% ...
$ ^{46}$H.H. Wills Physics Laboratory, University of Bristol, Bristol, United Kingdom\\
% ...
$ ^{51}$School of Physics and Astronomy, University of Glasgow, Glasgow, United Kingdom\\
$ ^{52}$Oliver Lodge Laboratory, University of Liverpool, Liverpool, United Kingdom\\
% ...
}
\end{flushleft}
%%%%%%%%%%%%%%%%%%%%%%%%%%%%%%%%%%%%%%%%%%